\documentclass[aps,twocolumn,pre,superscript,floatfix,superscriptaddress,10pt]{revtex4-2}

\usepackage{epsfig}
\usepackage{graphicx}
\usepackage{dcolumn}
\usepackage{bm}
\usepackage{amsfonts}
\usepackage{amsmath}
\usepackage{amssymb}
\usepackage{mathrsfs}
\usepackage{natbib}
\usepackage{multirow}
\usepackage{lineno}

\usepackage[colorlinks=true,%
            linkcolor=black,%
            urlcolor=black,%
            citecolor=black,%
            filecolor=blue,%
            bookmarksopen=true,%
            pdfauthor={M.Figueroa},%
            pdftitle={QR},%
            pdfsubject={QR_manuscript},%
            pdfpagemode=UseOutlines]{hyperref}
            
\usepackage[T1]{fontenc} 
\usepackage{lmodern}
\usepackage{xcolor}
\newcommand{\rev}[1]{#1}
\newcommand{\revb}[1]{#1}
\newcommand{\revc}[1]{#1}
\newcommand{\revd}[1]{#1}
\newcommand{\id}{\sigma_{0}}
\newcommand{\bnh}{\beta_{\rm NH}}

\newcommand{\bnhs}{\beta^2_{\rm NH}}

\begin{document}

\title{Thermodynamic signatures of spectral compression in weakly non-Hermitian Dirac fermions}

\author{Juan Pablo Esparza}
\thanks{These authors contributed equally to this work.}
\affiliation{Departamento de F\'isica, Universidad T\'ecnica Federico Santa Mar\'ia, Casilla 110, Valpara\'iso, Chile}
\affiliation{Instituto de F\'isica, Pontificia Universidad Cat\'olica de Valpara\'iso, Avenida Universidad 331, Curauma, Valpara\'iso, Chile}

\author{Francisco J. Peña}
\thanks{These authors contributed equally to this work.}
\affiliation{Facultad de Ingeniería, Universidad San Sebastián, Lago Panguipulli 1390, Puerto Montt, Chile}

\author{Patricio Vargas}
\affiliation{Departamento de F\'isica, Universidad T\'ecnica Federico Santa Mar\'ia, Casilla 110, Valpara\'iso, Chile}

\author{Vladimir Juri\v ci\'c}\thanks{Corresponding author:vladimir.juricic@usm.cl}
\affiliation{Departamento de F\'isica, Universidad T\'ecnica Federico Santa Mar\'ia, Casilla 110, Valpara\'iso, Chile}

\begin{abstract}
Non-Hermitian band descriptions capture how loss, gain, and environmental
coupling reshape quantum matter, yet most experimental probes remain wave
based or dynamical. Here we develop an equilibrium charge-response route to
spectral compression in Dirac materials. \rev{In the real-spectrum regime, invariance of the particle-number operator under the similarity transformation makes the quantum capacitance exactly that of the Hermitian partner, so the whole non-Hermitian content is carried by the Petermann factor.} In a minimal
non-reciprocal graphene model, hopping imbalance suppresses the Dirac
velocity, enhancing the low-energy density of states and the capacitance
\rev{as an exceptional point is approached}. At charge neutrality the capacitance stays linear in temperature, with its slope enhanced relative to the Hermitian value.
\rev{A perpendicular magnetic field recasts the same compression as a
collapse of the Landau-level ladder, drawing more levels into the thermal
window. In two dimensions, the capacitance enhancement relative to the
Hermitian partner coincides exactly with the Petermann factor, even though
the two have entirely different origins. This relation provides a
controlled starting point for extending equilibrium thermodynamic probes to
non-Hermitian electronic systems with interactions, disorder, and reservoir
coupling.}
\end{abstract}

\maketitle

\emph{Introduction.} Non-Hermitian (NH) physics has emerged as a powerful framework for open quantum systems, where loss, gain and dissipative coupling to the environment can qualitatively reshape spectra and response functions~\cite{Gong2018Topological,FoaTorres2020Perspective,Bergholtz2021Exceptional}. To date, experimental progress has been driven largely by wave-based platforms, including dissipative cold-atom arrays~\cite{Jin2017Topological,Li2020Topological,Liang2022Dynamic,Zhou2022Engineering,Cai2024NHSE,Zhao2025Two-dimensional}, photonic crystals~\cite{Zhang2024Topological}, optical waveguides~\cite{Zhou2018Optical} and topoelectrical circuits~\cite{Rafi-Ul-Islam2022Interfacial,Wu2022NH2nd,Liu2023Experimental,Ochkan2024NHTopology}. In parallel, NH Dirac materials~\cite{jurivcic2024yukawa,Murshed2024Quantum,Yu2024NHStrongly,Murshed2025Yukawa,leong2025arxiv,Roy-PRD-2025,Esparza2025EMAs} have opened a route towards quantum-material realizations, in which NH Dirac Hamiltonians arise as controlled low-energy descriptions generated by coupling to the environment or interactions~\cite{Rotter2009ANHH,Bergholtz2019NHWSM,Kozii2024NHTopological,Hamanaka2024NHTopology}. This brings exceptional spectral structures, including exceptional points (EPs), and NH topology directly into solid-state settings. A key question is how such NH spectral compression enters equilibrium thermodynamic response, and which part of the response can be distinguished from ordinary Hermitian velocity renormalization. We show that quantum capacitance provides a bulk thermodynamic probe of the weakly NH real-spectrum regime, while eigenstate non-orthogonality is separately captured by the Petermann factor.

\begin{figure}[t]
    \centering
    \includegraphics[width=.95\linewidth]{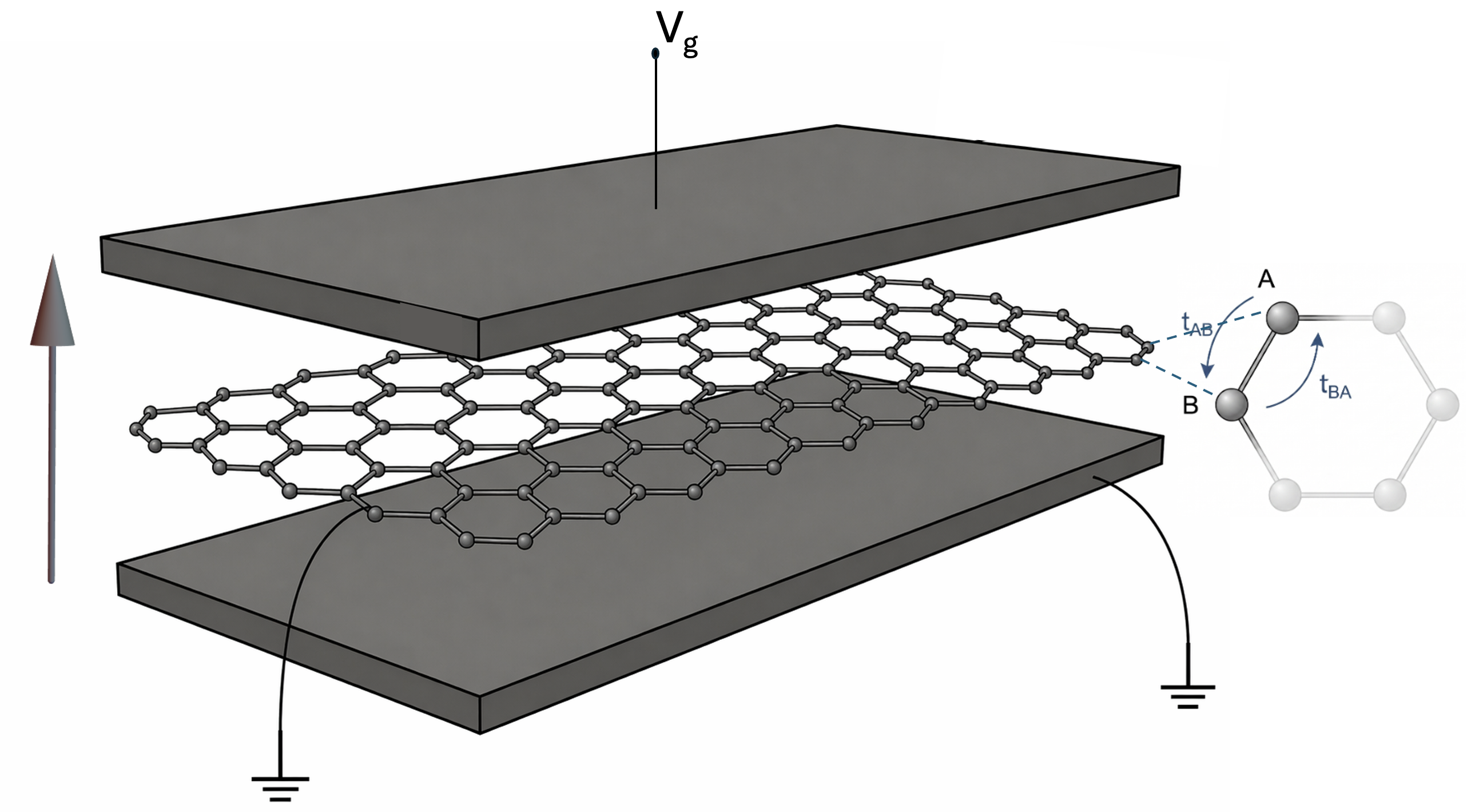}
    \caption{Proposed quantum-capacitance measurement setup. A two-dimensional Dirac sample is embedded in a capacitor and coupled to auxiliary reservoirs generating an effective real-spectrum NH Hamiltonian. The thermodynamic density of states (TDOS) is obtained from the quantum contribution to the capacitance using standard electrostatic techniques. A uniform magnetic field is applied perpendicular to the sample (left arrow). The inset illustrates asymmetric hopping between sublattices (A) and (B), with amplitudes 
($t_{AB} \neq t_{BA}$).}
    \label{fig:Exp_Setup}
\end{figure}

Quantum thermodynamics connects microscopic quantum degrees of freedom to macroscopic observables provided that an equilibrium (or stationary) ensemble exists \cite{Lindblad1976, Spohn1978, Alicki1979, Allahverdyan2005, Talkner2007, Esposito2009, Campisi2011, Deffner2011, Kosloff2013, Deffner2013, Kosloff2014, Gelbwaser2015, Goold2016, Vinjanampathy2016, Strasberg2017}. While generic NH dynamics can be non-unitary and exhibit complex spectra, pseudo-Hermitian Hamiltonians admit a consistent thermodynamic formulation: if the spectrum is real or appears in complex-conjugate pairs, the second law remains valid, and thermodynamic state functions can be defined~\cite{Gardas2016NHQT}. More broadly, for NH effective descriptions with a conserved notion of probability, via a  biorthogonal inner product or a controlled steady-state open-system realization, statistical mechanics remains well defined and yields macroscopic response functions~\cite{Bebiano2020Toward,Lima2021Information,Cao2023Statistical,Cao2025Information}. 

Among the experimentally accessible thermodynamic quantities of a sample, of particular interest in two-dimensional materials is the quantum capacitance $C_Q$, which arises as a \rev{series} contribution to the standard geometric contribution $C_G$ when a sample is placed inside a capacitor, \rev{$C_{\rm tot}^{-1}=C_G^{-1}+C_Q^{-1}$}~\cite{Luryi1988QCD}. This second contribution is related to the density of accessible states at any finite temperature $T$ at which the experiment is performed. Crucially, when such density is infinitesimally small, the effect on the measured capacitance is dominated by $C_Q$. As such, any deformation of the spectrum, and therefore of the allowed energies for a given temperature, would leave an imprint in the capacitance that can be experimentally measured.

\revb{Quantum capacitance is well established as a probe of electronic compressibility and density of states (DOS) in graphene and related two-dimensional systems~\cite{Luryi1988QCD,Xia2009Measurement,Ponomarenko2010DOS}. Capacitance also appears in NH sensing through distinct mechanisms, including $\mathcal{PT}$-symmetric electrical circuits~\cite{Chen2018PTX,Xiao2019PTSensing} and geometric contributions to quantum Hall admittance in NH Dirac systems~\cite{Fan2020Complex}. Here, instead, the quantum capacitance $C_Q=e^2(\partial n/\partial\mu)$ is the fermionic equilibrium charge compressibility generated by the spectrum and Fermi statistics. To the best of our knowledge, its use as a thermodynamic probe of NH spectral compression in a Dirac system has not been established previously.}

\revd{Our central result is a criterion separating spectrum-controlled thermodynamic response from eigenvector-sensitive NH observables. The criterion follows from the similarity transformation and its action on the sublattice operator basis; see Eq.~\eqref{eq:conjugation} and the accompanying discussion. In the zero-field two-dimensional Dirac regime, quantum capacitance and the Petermann factor exhibit the same divergence toward the EP, while encoding distinct physical content. Their locking is established below from the two-dimensional Dirac DOS.}

\emph{Model.} We consider a minimal NH extension of clean graphene where non-Hermiticity is introduced by means of a hopping imbalance between nearest neighbors~\cite{jurivcic2024yukawa}. The full tight-binding Hamiltonian then takes the form $H=\sum_\mathbf{k}\psi^\dagger_\mathbf{k}h_\mathbf{k}\psi_\mathbf{k}$, where the sublattice spinor $\psi_\mathbf{k}^\top=(a_\mathbf{k},b_\mathbf{k})$  and
\begin{equation}
    h_{\mathbf{k}}=-\begin{pmatrix}
        0 & t_{AB} f_\mathbf{k} \\ t_{BA} f^*_{\mathbf{k}} & 0
    \end{pmatrix}.
\label{eq:NH_graphene}
\end{equation}
Here $t_{AB}\in\mathbb{R}$ ($t_{BA}\in\mathbb{R}$) stands for the hopping amplitude from sites $A$ to $B$ ($B$ to $A$), and $f_\mathbf{k}=\sum_{\boldsymbol{\delta}}e^{-i\mathbf{k\cdot\boldsymbol{\delta}}}$, with $\boldsymbol{\delta}$ the corresponding vectors connecting  nearest-neighbor sites of the honeycomb lattice. The spectrum for this system is given by $\epsilon^{\pm}_\mathbf{k}=\pm\bar{t}\sqrt{1-\bnhs}|f_\mathbf{k}|$, where $\bar{t}=(t_{AB}+t_{BA})/2$, $\delta t=(t_{AB}-t_{BA})/2$ and we have introduced the dimensionless parameter $\bnh=\delta t/\bar{t}$ parameterizing the NH contribution. Notice that for $|\bnh|<1$ ($|\bnh|>1$), the spectrum is purely real (imaginary), with the EPs located precisely at $|\bnh|=1$. Clearly, for $t_{AB}=t_{BA}$, non-Hermiticity is suppressed as $\bnh=0$ and we recover the well-known Hermitian spectrum for clean graphene~\cite{gonzalez1993fullerenes,peres2006electronic,castro-neto2009electronic} with $\bar{t}\to t$; see Sec.~S2 of the Supplemental Material(SM)~\cite{SupplementalMaterial}. For the remainder of the work, we focus exclusively on the weakly NH regime, $|\bnh|<1$, where the single-particle spectrum remains real, and the equilibrium problem can be formulated in the biorthogonal basis corresponding to the effective NH Hamiltonian. Within the equilibrium pseudo-Hermitian formulation adopted here, the occupations are described by the usual Fermi--Dirac distribution, so the particle density retains the standard form $n=\int dE\,D(E)f(E-\mu)$ and the thermodynamic identity $C_Q=e^2(\partial n/\partial\mu)_{T,B,\bnh}$ remains valid throughout the weak-NH regime considered here~\cite{Gardas2016NHQT,Bebiano2020Toward,Cao2023Statistical}. In the present lattice model this can be shown explicitly: when $t_{AB}t_{BA}>0$ (equivalently $|\bnh|<1$), $h_{\mathbf k}$ is pseudo-Hermitian, since there exists a positive matrix $\eta={\rm diag}(\sqrt{t_{BA}/t_{AB}},\sqrt{t_{AB}/t_{BA}})$ such that $h_{\mathbf k}^{\dagger}=\eta\, h_{\mathbf k}\,\eta^{-1}$. Equivalently, the similarity transformation $\rho=\eta^{1/2}$ maps $h_{\mathbf k}$ onto the Hermitian partner $\rho\, h_{\mathbf k}\,\rho^{-1}=-\sqrt{t_{AB}t_{BA}}\begin{pmatrix}0&f_{\mathbf k}\\ f_{\mathbf k}^{*}&0\end{pmatrix}$, thereby establishing that the equilibrium ensemble is well defined throughout the weakly NH regime.

\revd{The same similarity transformation gives the operator criterion used below. For any observable $\mathcal{O}$,
$\langle\mathcal{O}\rangle_{\rm NH}
=\langle\rho\mathcal{O}\rho^{-1}\rangle_{\rm H}$,
where $\langle\cdots\rangle_{\rm NH}$ and $\langle\cdots\rangle_{\rm H}$ denote equilibrium averages in the NH and Hermitian-partner ensembles, respectively. Since the total particle-number operator satisfies $\rho N\rho^{-1}=N$, its equilibrium response maps directly onto that of the Hermitian partner.}

\revd{For the present two-band model, the action of $\rho$ is explicit on the complete sublattice operator basis. With
$\rho={\rm diag}(r,r^{-1})$, $r^{2}=\sqrt{t_{BA}/t_{AB}}$, one has
$\rho\,\id\,\rho^{-1}=\id$ and $\rho\,\sigma_z\,\rho^{-1}=\sigma_z$, whereas}
\begin{align}
\revc{\rho\,\sigma_x\,\rho^{-1}}&\revd{{}=\frac{\sigma_x-i\bnh\,\sigma_y}{\sqrt{1-\bnhs}},}\nonumber\\
\revc{\rho\,\sigma_y\,\rho^{-1}}&\revd{{}=\frac{\sigma_y+i\bnh\,\sigma_x}{\sqrt{1-\bnhs}}.}
\label{eq:conjugation}
\end{align}
\revd{Thus $N\propto\id$ is invariant and powers of $h_{\mathbf k}$ map algebraically to powers of its Hermitian partner. Because $\rho$ is momentum independent, the velocity operator also maps directly,
$\rho(\partial_{\mathbf k}h_{\mathbf k})\rho^{-1}
=\partial_{\mathbf k}h_{\mathbf k}^{(H)}$,
whereas $\sigma_x$ and $\sigma_y$ are mixed nontrivially by the similarity transformation.}

\revc{Accordingly, the grand potential in the real-spectrum regime reads,}
\begin{equation}
\Omega=-k_B T\sum_{\nu,\mathbf k}\ln\!\left[1+e^{-(E_{\nu,\mathbf k}-\mu)/k_B T}\right],
\end{equation}
leading to the usual expression for the TDOS via the Fermi-Dirac distribution. Equivalently, in a weakly open quasi-equilibrium realization, the charge response can be written as $C_Q=e^2\int d\omega\,[-\partial_\omega f(\omega)]\,{\rm Tr}\,A(\omega)/(2\pi)$, where $A(\omega)=i[G^R(\omega)-G^A(\omega)]$. The DOS formula used here is recovered in the narrow-linewidth limit; finite lifetime, disorder, noise, and reservoir broadening regularize the   singular behavior. More generally, Hilbert-space geometry enters NH observables such as Berry curvature and quantum-metric effects~\cite{DengCheng2025}, pseudo-Hermitian linear response~\cite{TetlingFistulEremin2022}, and oscillatory edge-state dynamics associated with non-orthogonality~\cite{KawabataSato2020}.

\begin{figure}[t!]
\centering
\includegraphics[width=\linewidth]{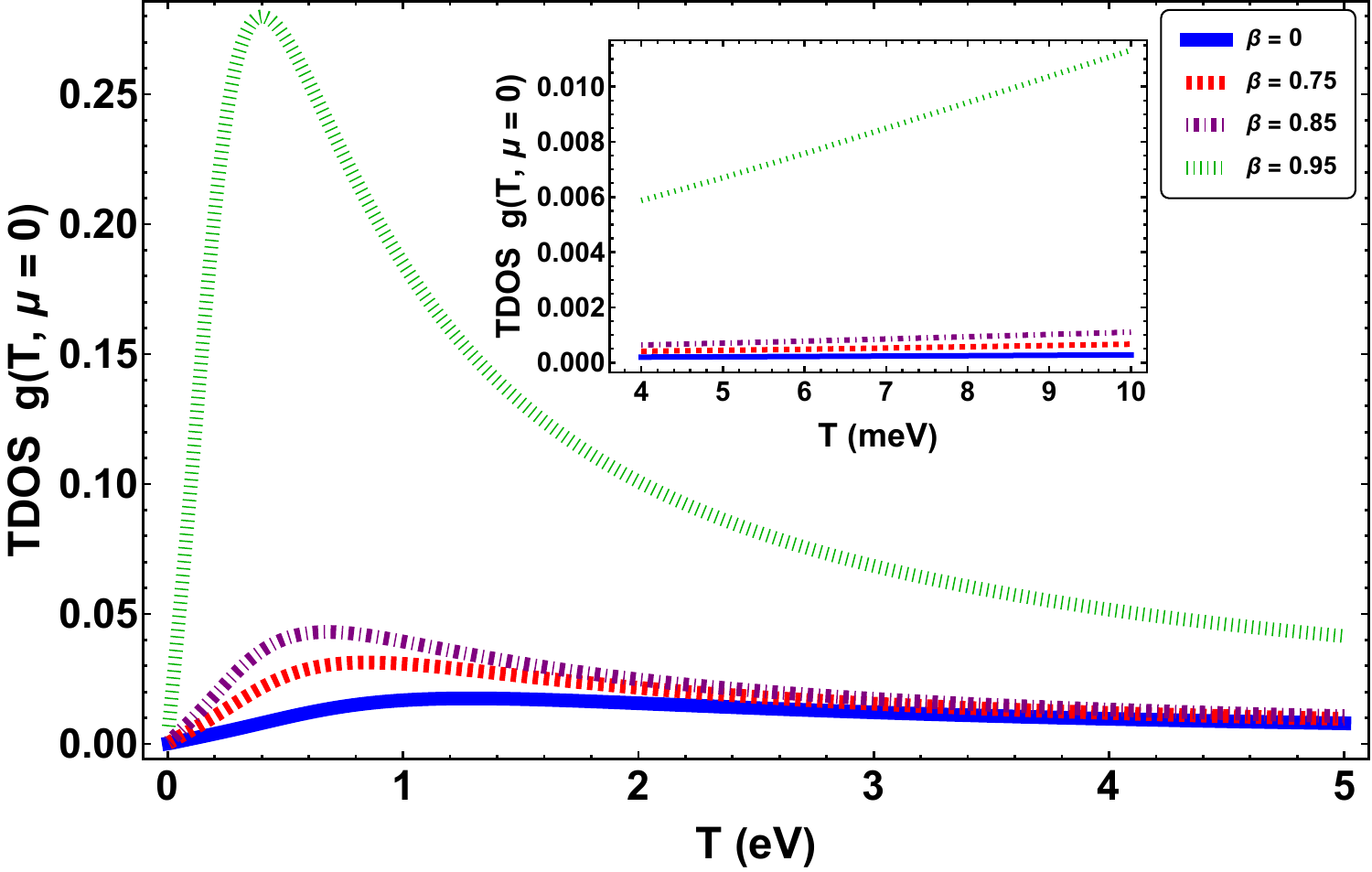}
\caption{Thermodynamic density of states $g(T,\mu{=}0)$ at half filling for pristine graphene ($\bnh=0$) and its NH deformation. The inset shows the universal low-$T$ Dirac regime, where $C_Q=e^2 g$ and $g\propto (1-\bnhs)^{-1}T$, implying a linear collapse of the inverse quantum capacitance on approaching the EP. The broader maximum at higher $T$ is a nonuniversal lattice-scale feature.}
\label{fig:TDOS_nonHermitian}
\end{figure}

\revd{The eigenvector-sensitive quantity used here is the Petermann factor, a standard measure of mode non-orthogonality~\cite{Petermann1979,Siegman1989I,Siegman1989II,ChengFanningSiegman1996,WangLaiYuanSuhVahala2020}. For the biorthogonally normalized Bloch eigenstates of Eq.~\eqref{eq:NH_graphene}, with normalized Hermitian-partner state $|u^{H}_{\mathbf{k},\lambda}\rangle$,}
\begin{align}
K_{\mathbf{k},\lambda}
&=\langle u^{R}_{\mathbf{k},\lambda}|u^{R}_{\mathbf{k},\lambda}\rangle\,
\langle u^{L}_{\mathbf{k},\lambda}|u^{L}_{\mathbf{k},\lambda}\rangle \nonumber\\
&=\langle u^{H}_{\mathbf{k},\lambda}|\rho^{-2}|u^{H}_{\mathbf{k},\lambda}\rangle\,
\langle u^{H}_{\mathbf{k},\lambda}|\rho^{2}|u^{H}_{\mathbf{k},\lambda}\rangle
\ge 1, \nonumber\\
&=\frac{1}{1-\bnhs}\qquad\text{for Eq.~\eqref{eq:NH_graphene}}.
\label{eq:Petermann_factor}
\end{align}
\revd{The inequality follows generally from Cauchy--Schwarz, while the final equality uses the equal sublattice weight of the eigenstates of the present model; hence $K$ is independent of $\mathbf{k}$ and $\lambda=\pm$ (see Sec.~S6 of the SM~\cite{SupplementalMaterial}). Thus $K$ directly measures mode non-orthogonality.}

In the proximity of the band-touching points $\mathbf{K}$ and $\mathbf{K'}$, this model exhibits an NH version of the two-dimensional Dirac equation~\cite{jurivcic2024yukawa}, and the low energy excitations of the system behave as Dirac fermions with a reduced Fermi velocity $v_F=v\sqrt{1-\bnhs}$. This depletion of the effective velocity has a direct and measurable impact in all the thermodynamic quantities through a modulation of the DOS, which scales as $\rho(E)\sim |E|/v^2_F$.

In the low-energy Dirac regime, the DOS per unit area is~\cite{jurivcic2024yukawa}
\begin{equation}
D_{\bnh}(E)=\frac{g_d}{2\pi(\hbar v_F)^2}|E|
=\frac{g_d}{2\pi(\hbar v)^2}\frac{|E|}{1-\bnhs},
\label{eq:D_beta_dirac}
\end{equation}
with $g_d$ the spin--valley degeneracy, which implies a universal approach to the EP in the spectrum,
\begin{equation}
D_{\bnh}(E)\propto (1-\bnhs)^{-1},
\, {\rm as}\quad |\bnh|\to 1^-.
\label{eq:DOS_EP_precursor}
\end{equation}
Observables controlled by the low-energy DOS therefore inherit the same enhancement.  \revd{Beyond transport signatures~\cite{jurivcic2024yukawa}, quantum capacitance provides an independent equilibrium probe of the same NH-induced spectral compression.}

\emph{Quantum capacitance.} To provide an explicit platform for the experimental exploration of our model, we focus on the quantum capacitance~\cite{Luryi1988QCD}, given its experimental relevance probing the properties of graphene-based structures, such as its spectral profile~\cite{Das2008Monitoring,Xia2009Measurement,Mivskovic2010Modeling,Zhan2015Quantum} and interaction phenomena~\cite{Iani2006Measurement,Yu2013Interaction}. A two-dimensional gas of Dirac fermions shows an enhanced TDOS at low temperatures, which induces a purely quantum contribution to the capacitance that can overcome the classical one~\cite{Luryi1988QCD}, and thus provide experimental access to the low-energy spectral properties of the material. \revc{Quantum capacitance is therefore sensitive to the suppression of the effective Fermi velocity and directly probes the associated spectral compression as $\bnh$ is tuned.}

In a gated two-dimensional device, the accumulation of charge is governed by the quantum capacitance per unit area,
\begin{equation}
C_{Q} = e^{2}\, g(T,B,\bnh,\mu),
\label{eq:CQ}
\end{equation}
which represents the electronic contribution to the total capacitance of the heterostructure, where $g(T,B,\bnh,\mu)\equiv (\partial n_e/\partial\mu)_{T,B,\bnh}$ is the TDOS, which coincides with the local DOS in the limit $T\to0$ (see Sec.~S1 of the SM~\cite{SupplementalMaterial}). 
A small shift in chemical potential changes the carrier density according to
\begin{equation}
\color{black}
dn_{e}=g\,d\mu=\frac{C_Q}{e^2}\,d\mu,
\label{eq:dNe}
\end{equation}
while the applied gate voltage is related by
$e\,dn_e=C_{\rm tot}\,dV_g$, with
$C_{\rm tot}^{-1}=C_G^{-1}+C_Q^{-1}$.

Moreover, in the Dirac regime  the TDOS can be evaluated analytically from Eq.~\eqref{eq:CQ}. Using Eq.~\eqref{eq:D_beta_dirac} together with the thermal-window representation
$g(T,B,\mu,\bnh)=\int dE\,D_{\bnh}(E)\bigl(-\partial f/\partial E\bigr)$, at $B=0$ one finds (see Sec.~S1 of the SM~\cite{SupplementalMaterial} for details)
\begin{equation}
g(T,B=0,\mu,\bnh)=
\frac{g_d k_B T}{\pi(\hbar v)^2}\,
\frac{\ln\!\left[2\cosh\!\left(\frac{\mu}{2k_B T}\right)\right]}
{1-\bnhs}.
\label{eq:g_exact_dirac}
\end{equation}
The quantum capacitance thus takes the  scaling form
\begin{align}
&C_Q(T,\mu,\bnh)=e^2 g(T,B=0,\mu,\bnh)\nonumber\\
&=\frac{1}{1-\bnhs}\,C_Q^{(H)}(T,\mu),
\label{eq:CQ_exact_scaling}
\end{align}
where $C_Q^{(H)}$ denotes the Hermitian Dirac result computed with the bare velocity $v$ and $B=0$. 
\revd{Equation~\eqref{eq:CQ_exact_scaling} makes the spectrum-controlled capacitance enhancement explicit. Equations~\eqref{eq:CQ_exact_scaling} and \eqref{eq:Petermann_factor} therefore imply the exact relation
${C_Q}/{C_Q^{(H)}}=K.$
The locking is specific to two dimensions. Since
$v_F/v=\sqrt{1-\bnhs}=K^{-1/2}$ and
$D_d\propto v_F^{-d}$, the Dirac DOS obeys
$D_d/D_d^{(H)}=K^{d/2}$, yielding
$C_Q/C_Q^{(H)}=K$ for $d=2$.}
Two limits are especially transparent in the zero-field ($B=0$) limit:
\begin{align}
&g(T,0,\bnh)=
\frac{g_d \ln 2}{\pi(\hbar v)^2}\,
\frac{k_B T}{1-\bnhs},\quad{\rm and}\quad\nonumber\\
&g(0,\mu,\bnh)=
\frac{g_d |\mu|}{2\pi(\hbar v)^2}\,
\frac{1}{1-\bnhs}.
\label{eq:g_limits_dirac}
\end{align}
Equation~\eqref{eq:g_limits_dirac} identifies a clean thermodynamic signature of the EP: the NH deformation drives a universal divergence of the compressibility and quantum capacitance, while the inverse response softens linearly, $g^{-1}\sim C_Q^{-1}\propto 1-\bnhs$, as $|\bnh|\to 1^-$.  Moreover, the measured capacitance obeys $C_{\rm tot}^{-1}=C_G^{-1}+C_Q^{-1}$; hence a large $C_Q$ produces an anomalously rapid crossover of $C_{\rm tot}$ toward the geometric capacitance $C_G$, rather than an unbounded measured capacitance. \revd{The Dirac scaling itself requires}
\rev{$\max(|\mu|,k_BT,\Gamma)\ll E_{\Lambda}(\bnh)\sim\hbar v_F\Lambda$.
Because $v_F=v\sqrt{1-\bnhs}$, the effective Dirac window shrinks on approaching the EP, so at fixed finite $\mu$, $T$, or linewidth $\Gamma$ the scaling eventually breaks down and the formal $(1-\bnhs)^{-1}$ divergence is cut off before $|\bnh|=1$.}
{\revc{Hence, monitoring $C_Q$ as $\bnh$ is tuned provides direct access to the associated spectral compression.} The key point is that, in the Dirac regime, the low-energy TDOS obeys the analytic scaling law in Eq.~\eqref{eq:g_exact_dirac}, yielding a collapse of $(1-\bnhs)g/T$ onto a universal function of $\mu/k_B T$. At half filling, $C_Q(T,\mu{=}0)$ therefore grows linearly with temperature but with a prefactor diverging as $(1-\bnhs)^{-1}$, while the inverse quantum capacitance collapses linearly as the EP is approached. Figure~\ref{fig:TDOS_nonHermitian} may thus be interpreted as the finite-temperature manifestation of the universal equilibrium approach to the EP in the compressibility.}
To illustrate this, Fig.~\ref{fig:TDOS_nonHermitian} shows the TDOS at fixed chemical potential $\mu=0$ for representative values of the NH parameter. The low-temperature sector highlighted in the inset is the universal Dirac regime governed by Eq.~\eqref{eq:g_limits_dirac}: there the TDOS is linear in $T$ at half filling, with a prefactor enhanced by $(1-\bnhs)^{-1}$. By contrast, the broad maximum at higher temperature reflects nonuniversal lattice-scale contribution to the DOS obtained from the tight-binding model, as shown in Sec.~S2 of the SM~\cite{SupplementalMaterial}.

\emph{Magnetic field effects.} Motivated by the zero-field results, we next explore the behavior of the model in the presence of a perpendicular magnetic field $B$. In the low-energy Dirac description the spectrum retains the relativistic graphene form, with the NH deformation entering only through the reduced velocity $v_F=v\sqrt{1-\bnhs}$. The LLs are therefore (see Sec.~S3 of the SM~\cite{SupplementalMaterial} for details)
\begin{equation}
    {\epsilon_{n}^{\pm}(B,\bnh)=\pm v\sqrt{2e\hbar B\,(1-\bnhs)\,n},\qquad n=0,1,2,\ldots,}
\label{eq:NH_LLs}
\end{equation}
so that the $n=0$ level remains pinned at zero energy while the first nonzero level is $\epsilon_1=v\sqrt{2e\hbar B\,(1-\bnhs)}$, which is the convention used hereafter. As in Hermitian graphene~\cite{mcclure1956diamagnetism,zheng2002hall,Lukose2007Novel,Peres2007Algebraic,jiang2007infrared,li2007observation,castro-neto2009electronic,koshino2011landau,luican2011quantized}, the quantized ladder leaves a characteristic imprint in oscillatory and thermodynamic observables. Here the role of non-Hermiticity is to compress the entire ladder by the universal factor $\sqrt{1-\bnhs}$, thereby shifting all finite-energy LLs in a controlled way. Throughout, spin is included via the spin-valley degeneracy $g_d=4$, while Zeeman splitting is neglected as it is small compared to the LL spacing in the field range considered. This isolates the effect of the NH parameter $\bnh$ on the orbital dynamics and thermodynamic response, as detailed in Sec.~S4 of the SM~\cite{SupplementalMaterial}.

\emph{Effects of magnetic field on the quantum capacitance.} Following the same experimental protocol, we propose to extract the spectral properties of the NH Dirac model from the quantum capacitance of a two-dimensional sample in a perpendicular magnetic field. In this case the TDOS develops sharp peaks whenever the chemical potential crosses a LL. Using  Eq.~\eqref{eq:NH_LLs}, the finite-temperature TDOS is
\begin{equation}
{
g=\frac{\mathcal N_B}{4k_BT}
\sum_{n=0}^{\infty}\sum_{s=\pm}
w_{n,s}\,
{\rm sech}^{2}\!\left[\frac{s\,\epsilon_n(B,\bnh)-\mu}{2k_BT}\right],}
\label{eq:TDOS_LL}
\end{equation}
where $w_{0,\pm}=\tfrac12$,  $w_{n\ge1,\pm}=1$,
and  $\epsilon_n(B,\bnh)=v\sqrt{2e\hbar B\,(1-\bnhs)\,n}$ for $n=1,2,\ldots$, the first term is the isolated $n=0$ LL pinned at zero energy, and $\mathcal{N}_B=g_d eB/(2\pi\hbar)$ is the total LL degeneracy per unit area, including spin and valley (see Sec.~S5 of the SM~\cite{SupplementalMaterial}). 

\begin{figure}[t]
\centering
\includegraphics[width=\linewidth]{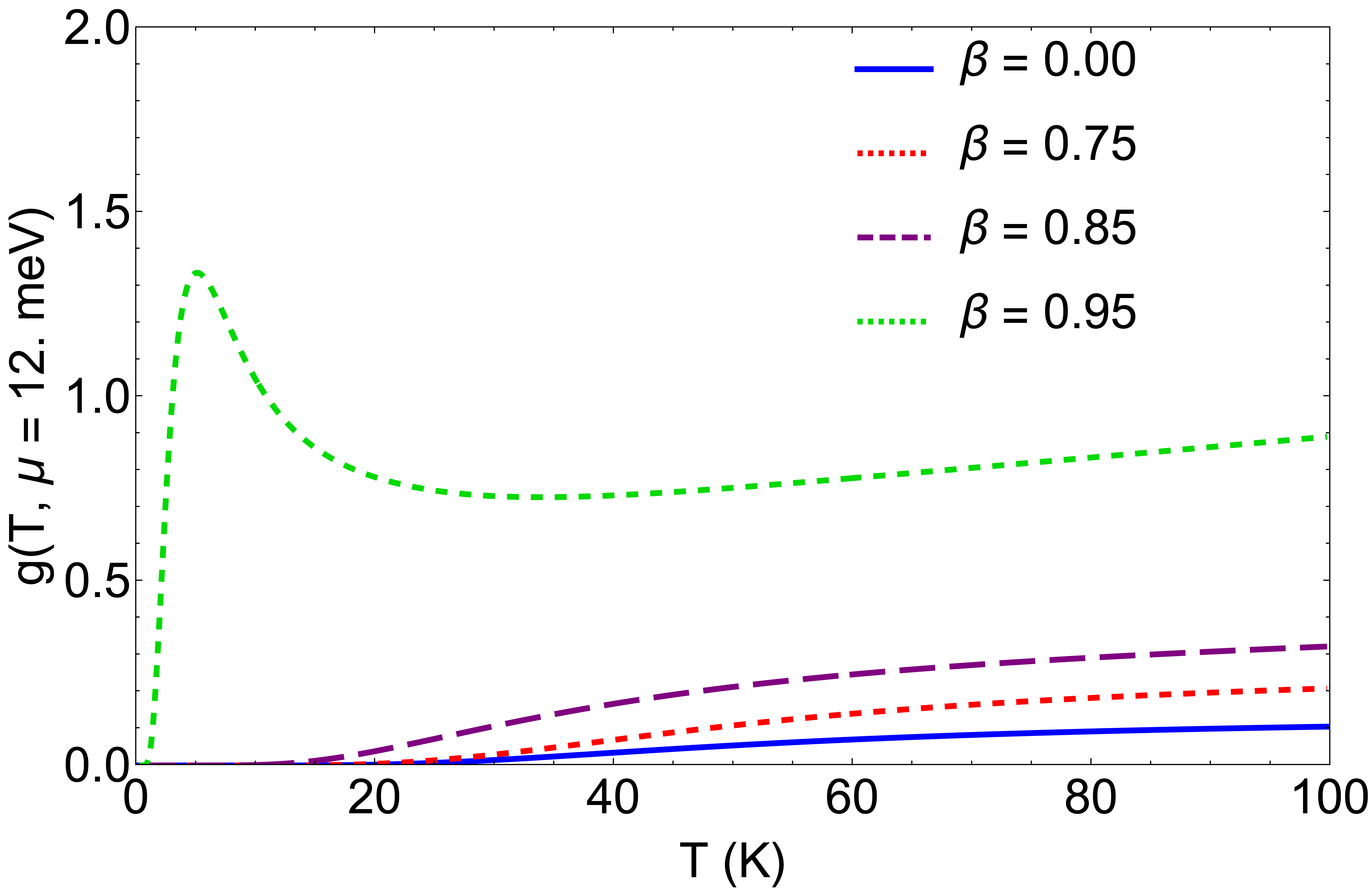}
\caption{Thermodynamic density of states $g(T,\mu)$ at $B=1\,$T and $\mu=12\,$meV for several non-Hermitian deformations $\bnh$. Since $C_Q=e^2 g$, \revc{the same low-$T$ enhancement appears in the quantum-capacitance contribution to the measured response.} Among the values of $\bnh$ shown in the figure, the strongest enhancement occurs at $\bnh=0.95$, where $\mu$ lies closest to the $n=1$ LL; for smaller $\bnh$, the level shifts away and the temperature dependence becomes weaker.}

\label{fig:TDOS_T_mu12}
\end{figure}

{
The finite-field response can also be stated in a compact analytic form. Since the NH deformation only rescales the LL spectrum through the reduced velocity, with the relevant scale behaving as 
\begin{equation}
E_{\rm LL}(B,\bnh)\propto v\sqrt{1-\bnhs}\,\sqrt{eB}.
\label{eq:ELL_scale}
\end{equation}
It follows that the number of LLs contributing within the thermal window around the chemical potential scales as
\begin{equation}
N_T \sim \left(\frac{k_B T}{E_{\rm LL}(B,\bnh)}\right)^2
\propto \frac{1}{1-\bnhs}.
\label{eq:NT_scaling}
\end{equation}
Hence, as $|\bnh|\to 1^-$, the finite-field enhancement of the TDOS is governed by LL crowding: more and more quantized levels are compressed into the thermal window, and the discrete TDOS crosses over to the same $(1-\bnhs)^{-1}$ scaling found in zero field once $k_B T\gtrsim E_{\rm LL}(B,\bnh)$. This finite-field response is therefore the Landau-quantized manifestation of the same spectral compression that controls the zero-field capacitance, rather than a separate eigenstate-sensitive diagnostic. \rev{Likewise, the Landau-level crowding is a low-energy Dirac asymptote, since the finite lattice bandwidth bounds the number of available levels and therefore cuts off the accumulation.}
}{Figure~\ref{fig:TDOS_T_mu12} compares the finite-field TDOS for different values of $\bnh$, with the chemical potential chosen near the first nonzero LL for $\bnh=0.95$. The resulting low-temperature maximum therefore corresponds to the $n=1$ level [see Eq.~\eqref{eq:NH_LLs}], while for smaller $\bnh$ the same level moves away from the chosen chemical potential and the response becomes smoother.}

\begin{figure}[t!]
\centering
\includegraphics[width=\linewidth]{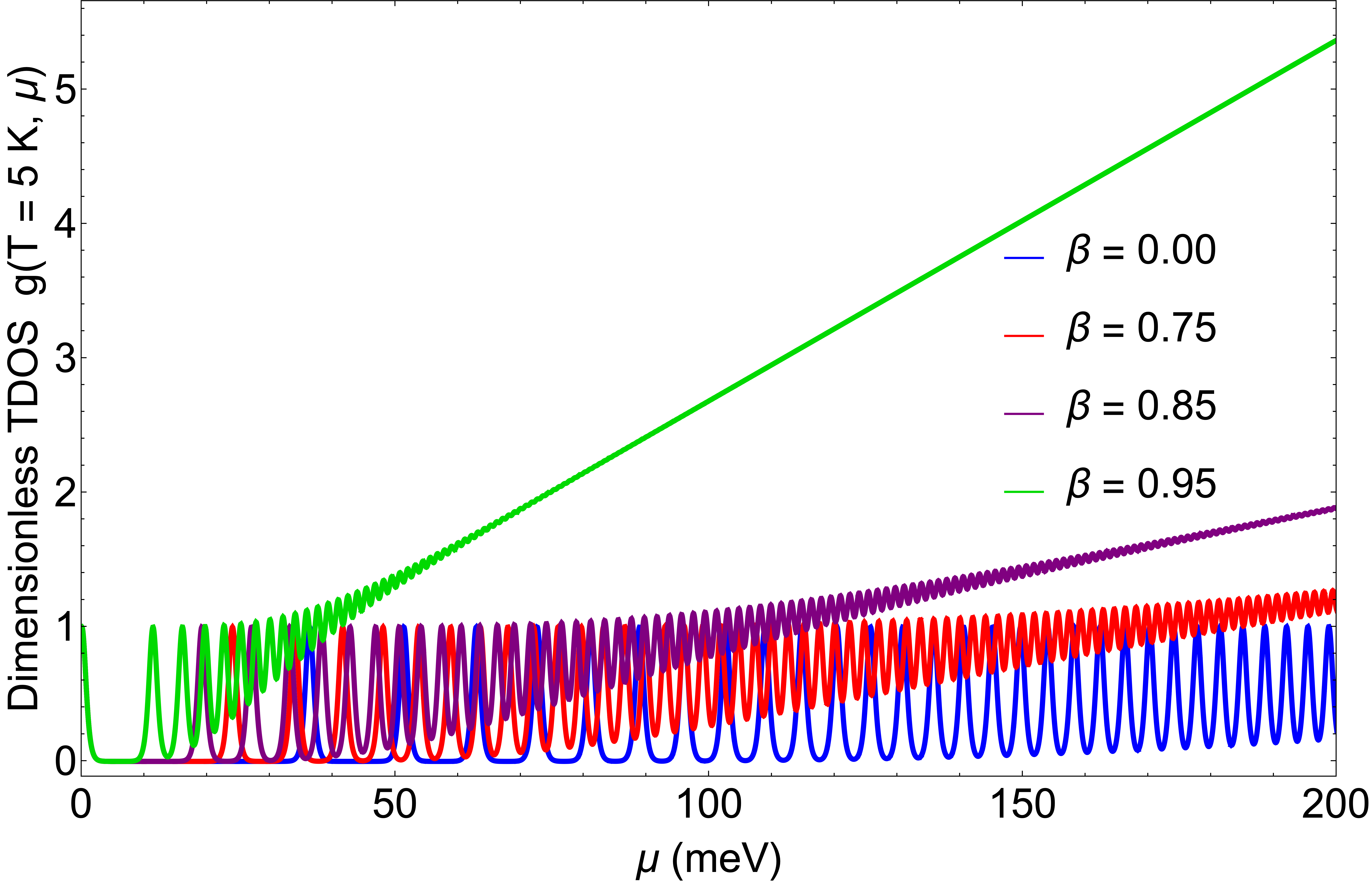}
\caption{Dimensionless thermodynamic density of states $g(T{=}5\,\mathrm{K},\mu)$ of monolayer graphene at $B=1\,$T for several NH deformations $\bnh$. Peaks occur when the chemical potential crosses the Landau levels. As $\bnh$ increases, the Landau ladder is compressed by $\sqrt{1-\bnhs}$, producing a denser oscillatory pattern. Since $C_Q=e^2 g$, the same trend provides a direct quantum-capacitance signature of non-Hermitian Landau-level compression. The gradual reduction of peak heights at larger $\mu$ reflects thermal smearing of higher Landau levels.}
\label{fig:TDOS_landau_beta}
\end{figure}

{Alternatively, a measurement of the TDOS at fixed temperature and variable chemical potential directly resolves the redistribution of LL peaks for any finite value of $\bnh$ [see Eq.~\eqref{eq:NH_LLs}]. As the NH contribution grows, the quantized levels become closer together for all $n\ge 1$, so the oscillatory pattern in $g(T,\mu)$ becomes progressively denser. Figure~\ref{fig:TDOS_landau_beta} illustrates this compression of the LL ladder of states and provides a second thermodynamic signature of effective non-Hermiticity in a magnetic field.}

\emph{Experimental relevance.} Quantum thermodynamics of NH systems has been shown to be theoretically consistent, allowing for recent studies and proposals for thermodynamic measurements in NH setups, including lattice structures~\cite{Pyrialakos2022Thermalization,Munoz-Arboleda2024Thermodynamics}, Josephson junctions~\cite{Pino2025Thermodunamics}, and relativistic models~\cite{Felski2023Thermodynamic}. In this context, our work proposes a charge-response protocol for detecting the thermodynamic consequences of NH spectral compression. For electronic Dirac materials, quantum capacitance is the natural observable; for engineered photonic, polaritonic, topoelectrical, or cold-atom Dirac simulators, the corresponding measurement would be a spectral density, integrated DOS, or analog compressibility.

The model is formulated as a minimal effective Dirac description of a real-spectrum NH regime. 
At the effective level, one may envision a gated Dirac layer, such as graphene or a surface Dirac state, 
coupled to auxiliary reservoirs whose low-energy self-energy suppresses the effective velocity while maintaining an approximately real quasiparticle dispersion with narrow linewidths. 
In such a setting, the parameter $\bnh$ can be inferred from equilibrium data through the collapse 
of $(1-\bnhs)g/T$ and through LL crowding at fixed $B$, together with independent characterization 
of the microscopic mechanism responsible for the velocity suppression. 
Reservoir-induced linewidths, disorder, and other weak broadening mechanisms can 
be incorporated through the spectral-function expression above, where they set the physical 
regularization scale for the ideal scaling. 
The Petermann factor, which is experimentally relevant in NH wave 
systems~\cite{ChengFanningSiegman1996,WangLaiYuanSuhVahala2020}, provides a complementary 
benchmark of mode non-orthogonality together with the thermodynamic capacitance response. 
The spectral-function formulation applies to the narrow-linewidth quasi-equilibrium 
regime with Fermi--Dirac occupations; away from this limit, the charge response is governed by 
the lesser (Keldysh) Green function~\cite{HaugJauho2008}. A microscopic electronic realization of 
$t_{AB}\neq t_{BA}$ preserving this quasi-equilibrium condition remains a challenge for future work.

\emph{Discussion and Outlook.} We have formulated an equilibrium thermodynamic route for probing spectral compression in a weakly NH Dirac model. In the low-energy Dirac regime, the TDOS and quantum capacitance scale as $(1-\bnhs)^{-1}$ as the EP is approached, while the same reduction of the Dirac velocity compresses the Landau-level spectrum at finite magnetic field.

\revb{The central physical result is the exact separation between thermodynamic and eigenvector-sensitive observables. Quantum capacitance provides the equilibrium readout of spectral compression, whereas the Petermann factor probes the accompanying mode non-orthogonality. In two dimensions, the two quantities exhibit the same divergence toward the EP while retaining distinct physical content. Their comparison therefore provides a controlled benchmark for weakly NH Dirac systems and a basis for extending the analysis to more microscopic electronic settings with finite lifetimes, interactions, and reservoir coupling.}

\emph{Data availability.} The data are available from the corresponding author upon reasonable request.

\emph{Acknowledgments.} This work is supported by Fondecyt (Chile) Grants  No. 1230933 (V.J.), 1240582 (P.V.) and 1250173 (F.J.P.). J.P.E. acknowledges support from Agencia Nacional de Investigación y Desarrollo (ANID) – Scholarship Program through the Doctorado Nacional Grant No. 2024-21240412. P.V. also acknowledges support from CEDENNA under grant CIA No.~250002.

\bibliography{Bibliography}

\end{document}